\documentclass{article}

\usepackage{titling}
\usepackage{arxiv_template}

\usepackage[utf8]{inputenc} 
\usepackage[T1]{fontenc}    
\usepackage{hyperref}       
\usepackage{url}            
\usepackage{booktabs}       
\usepackage{amsfonts}       
\usepackage{nicefrac}       
\usepackage{microtype}      
\usepackage{graphicx}
\usepackage{natbib}
\usepackage{doi}
\usepackage{xcolor}

\usepackage{fvextra}
\usepackage{wrapfig}
\usepackage{subcaption}

\usepackage{amsmath}

\title{\textit{ChartGen}: Scaling Chart Understanding Via Code-Guided Synthetic Chart Generation}

\author{
    \begin{tabular}{c}
    \textbf{Jovana Kondic$^{1}$, Pengyuan Li$^{3}$, Dhiraj Joshi$^{3}$, Zexue He$^{2}$, Shafiq Abedin$^{3}$, Jennifer Sun$^{1}$,} \\
    \textbf{Ben Wiesel$^{3}$, Eli Schwartz$^{3}$, Ahmed Nassar$^{3}$, Bo Wu$^{2,3}$, Assaf Arbelle$^{3}$,}\\
    \textbf{Aude Oliva$^{1,2}$, Dan Gutfreund$^{2,3}$, Leonid Karlinsky$^{2,3}$, Rogerio Feris$^{2,3}$}\\[6pt]
    \end{tabular}\\
    {\normalsize $^1$MIT \quad $^2$MIT-IBM Watson AI Labs} \quad $^3$IBM Research\\[4pt]
    {\normalsize \texttt{jkondic@mit.edu}}
}

\date{}

\begin{document}
\maketitle

\begin{abstract}
	\textit{Chart-to-code reconstruction} -- the task of recovering executable plotting scripts from chart images -- provides important insights into a model's ability to ground data visualizations in precise, machine-readable form. Yet many existing multimodal benchmarks largely focus primarily on answering questions about charts or summarizing them. To bridge this gap, we present \textit{ChartGen}, a fully-automated pipeline for code-guided synthetic chart generation. Starting from seed chart images, ChartGen (i) prompts a vision-language model (VLM) to reconstruct each image into a python script, and (ii) iteratively augments that script with a code-oriented large language model (LLM). Using ChartGen, we create 222.5\,K unique chart-image code pairs from 13\,K seed chart images, and present an open-source synthetic chart dataset covering 27 chart types, 11 plotting libraries, and multiple data modalities (image, code, text, CSV, DocTags). From this corpus, we curate a held-out chart-to-code evaluation subset of 4.3\,K chart image-code pairs, and evaluate six open-weight VLMs (3\,B - 26\,B parameters), highlighting substantial room for progress. We release the pipeline, prompts, and the dataset to help accelerate efforts towards robust chart understanding and vision-conditioned code generation: \url{https://github.com/SD122025/ChartGen/}.
\end{abstract}

\keywords{dataset and benchmark \and VLMs \and LLMs \and chart reasoning \and code generation \and data visualization}

\section{Introduction}

Visual representations of data are integral to effectively communicating insights and context in domains ranging from scientific research to business analytics. One of the most widely used methods for data visualization in data science is the Python plotting ecosystem, particularly libraries such as \texttt{matplotlib}, \texttt{seaborn}, and \texttt{plotly}. Large language models (LLMs), especially those trained on code-centric tasks, have demonstrated promising capabilities in generating plotting scripts for a variety of visualization needs \citep{chen2021evaluating}.
However, the reverse task -- i.e., \textit{chart-to-code reconstruction}, which involves extracting plotting code from a chart image -- has received comparatively little attention in vision-language research. 

Chart-to-code reconstruction is critical both for evaluating the chart parsing skills of vision-language models (VLMs) and for facilitating downstream applications such as automated chart analysis, data extraction, and advanced chart reasoning. By converting chart pixels into code, one effectively transforms the complex, high-dimensional pixel domain into a structured, reusable format that can be easily manipulated by LLMs and other computational tools.

\begin{figure}
    \centering
    \includegraphics[width=1.0\linewidth]{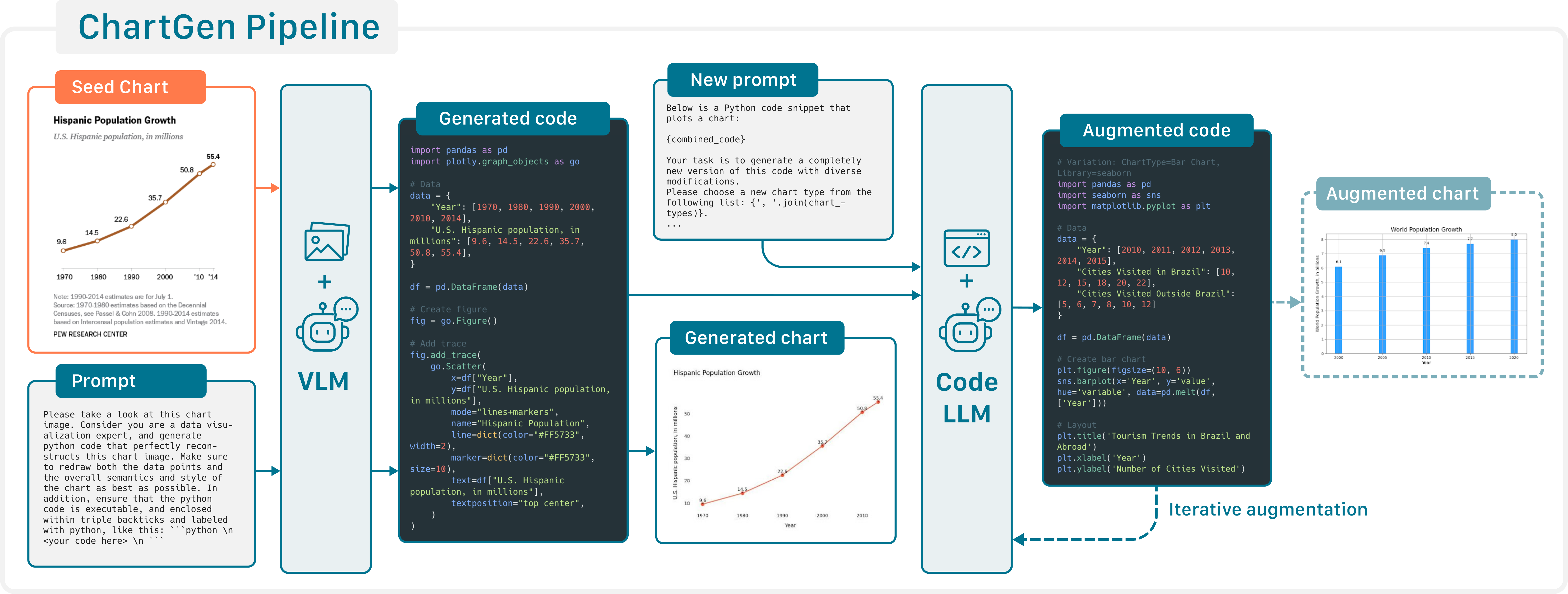}
    \caption{An overview of ChartGen: an automated pipeline for code-guided synthetic chart generation. First, a seed chart image is passed to a vision-language model for chart redrawing -- translating the image into executable plotting code. Then, the generated code is passed to a large language model, and iteratively augmented to collect diverse outputs.}
    \label{fig:pipeline}
\end{figure}

In this paper, we propose \textit{ChartGen}: a scalable, automated two-stage pipeline for generating large-scale, multimodal synthetic chart datasets. Building on a small seed set of chart images, ChartGen dramatically expands the size and diversity of chart-image/code pairs by leveraging two primary components: 1) VLM-based chart image redrawing, in which a vision-language model is used to convert chart images into Python plotting scripts that approximately reconstructs them, and 2) LLM-based chart code augmentation, where a code-focused large language model refines and diversifies the VLM-generated Python plotting code, ultimately producing new plotting scripts and charts of varied types, styles, and complexities.

Using ChartGen, we construct a synthetic dataset comprising 222.5\,K chart image-code pairs spanning 27 chart types, and 11 plotting libraries, as well as a variety of topics and layouts. In addition, we provide supporting chart data including underlying tabular data, DocTags \citep{smoldocling}, natural language summaries, and question-answer (QA) pairs.

From this corpus, we construct an evaluation subset of 4.3\,K chart image-code pairs, and develop an automatic protocol for evaluating chart-to-code reconstruction performance that scores data fidelity, code similarity, and rendered image similarity using GPT-4o as a reference judge.
Experiments on six open-weight VLMs ranging from 3 B to 26 B parameters show that, even when a large fraction of generated scripts executes without error, models still fall short of faithfully reconstructing charts (the best model attains only 0.58/1 data fidelity, 5.83/10 on code similarity and 7.48/10 image similarity), indicating substantial room for improvement in chart-to-code reconstruction and vision-conditioned code generation more broadly.

Our work makes three contributions:
\begin{enumerate}
    \item \textbf{ChartGen pipeline.}  
    We introduce a fully automated, two–stage synthetic chart generation pipeline that first reconstructs real‐world chart images into code with a vision–language model and then iteratively augments the resulting code with a code-centric LLM. ChartGen can be run with \emph{any} collection of seed charts (even those lacking source code), and increases the diversity and scale of the seed collection by orders of magnitude -- bridging the gap between small, domain-specific chart corpora and the massive scale required for modern multimodal training.

    \item \textbf{ChartGen-200K dataset.}  
    Running the pipeline on 13\,K unique seed chart images from ChartQA \citep{masry2022chartqa} produces a synthetic corpus of 200\,K image–code pairs that spans 27 chart types, 11 Python plotting libraries, and diverse stylistic variations. Each pair is accompanied by the underlying tabular data (CSV), a DocTags representation, a concise text summary, and automatically-generated QA pairs, resulting in a comprehensive public resource for chart understanding research.  Compared with existing chart-to-code collections,  
    ChartGen is up to two orders of magnitude larger and supports more chart types and plotting back-ends.

    \item \textbf{Chart redrawing evaluation.}  
    From the corpus we curate a held-out evaluation set of 4.3\,K chart image-code pairs covering a variety of chart types and libraries, and we propose a GPT-4o–based chart redrawing evaluation protocol that scores both the generated code and the resulting rendered image.  
\end{enumerate} 

By releasing the ChartGen code, dataset, and benchmark under an open license (\href{https://creativecommons.org/licenses/by/4.0/deed.en}{CC-BY-4.0 License}), we aim to help catalyze the progress toward robust automated chart understanding.

\section{The ChartGen Pipeline}
\label{subsec:data_gen}

In this section, we introduce \textit{ChartGen}, a fully-automated pipeline designed for code-guided synthetic chart generation. ChartGen transforms an initial seed dataset of chart images into structured code representations, subsequently scaling and diversifying the dataset through iterative augmentations. The pipeline consists of two stages: 1) VLM-based chart-to-code reconstruction, and 2) LLM-based chart code augmentation, as outlined in Figure~\ref{fig:pipeline}. 

\begin{figure}[htbp]
  \centering
  \includegraphics[width=1.0\linewidth]{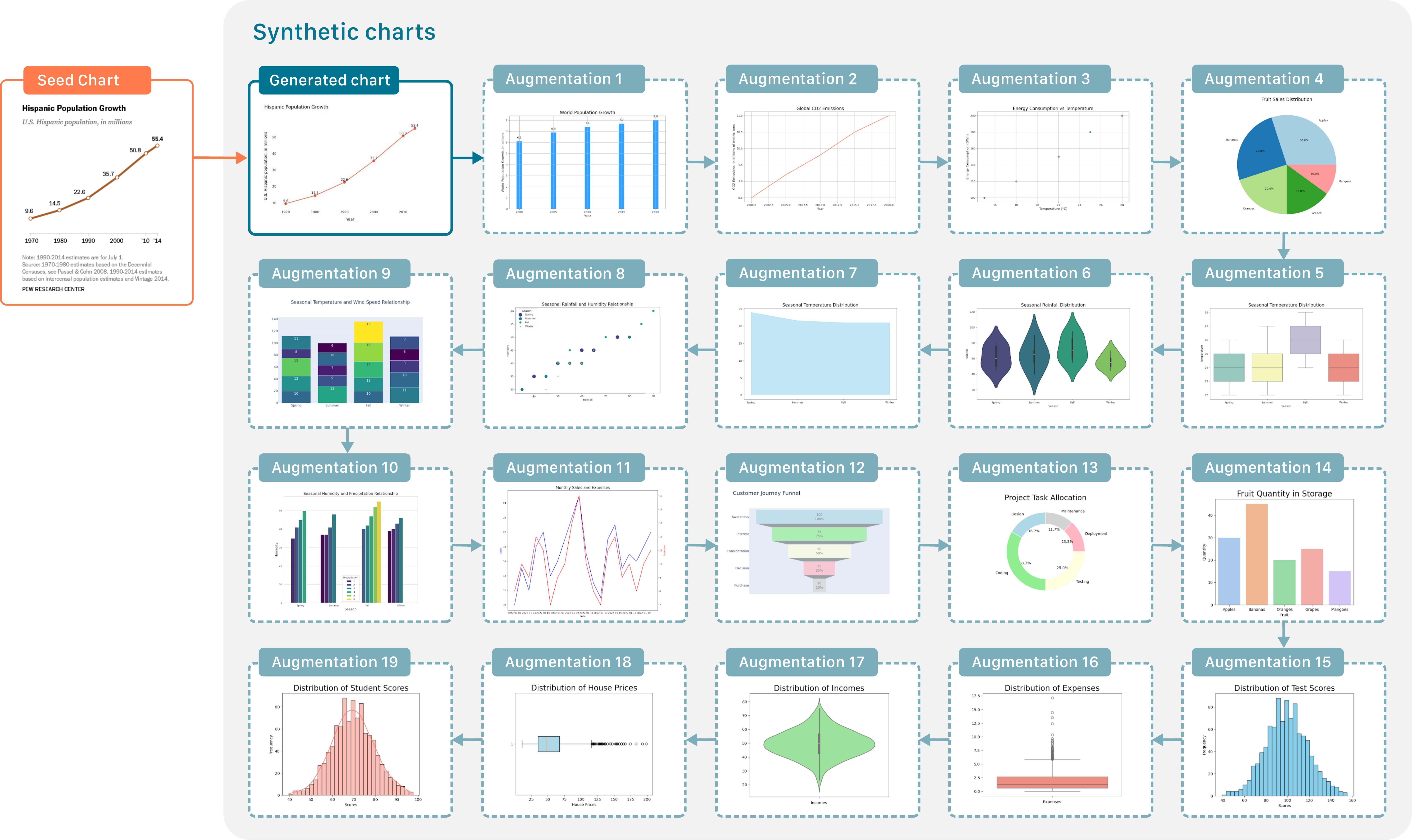} 
  \caption{An illustration of synthetic chart images generated from a single seed chart using the ChartGen pipeline. Initially, a \textit{seed chart} image is translated into an approximately matching plotting code, whose execution renders the first \textit{generated chart}. The code is then iteratively augmented, producing diverse variations in chart types, styles, and topics, as illustrated by the subsequent \textit{augmentations}}
  \label{fig:augmentation process}
\end{figure}

\subsection{VLM-based Chart-to-Code Reconstruction}  

Given a seed set of chart images, we prompt a vision-language model, \texttt{phi-3.5-vision-instruct}, to produce Python plotting code that reconstructs each chart image. For our experiments, we select a seed set of 13\,K unique chart images from the ChartQA dataset \citep{masry2022chartqa}, which contains real-world charts sourced from diverse domains and is widely used in the chart-understanding literature. Notably, ChartGen is flexible enough to accommodate any seed dataset—even when corresponding source code is unavailable, or when the charts were originally created using non-Python plotting methods (as is the case with ChartQA).

After processing the entire seed set, we obtain approximately 13\,K plotting scripts corresponding to the original 13\,K chart images. These scripts do not necessarily match the input charts exactly; rather, their primary role is to provide an initial structured code-based representation of the chart content.

\subsection{LLM-based Chart Code Augmentation} 
Next, leveraging the generative capabilities of a code-focused large language model (4-bit quantized \texttt{Codestral-22B-v0.1}), we iteratively augment the plotting scripts generated in the first stage.
The augmentation process involves multiple sequential transformations ($\sim$ 20 iterations) for each initial plotting script. These transformations are systematically guided by tailored prompts designed to produce visually and structurally distinct charts through variation of plotting libraries, chart types, data distributions, thematic topics, and stylistic attributes. Crucially, instead of altering the rendered image as typical data-augmentation techniques do, we transform the underlying plotting code itself -- a richer, fully structured representation that preserves the chart’s complete semantics

Execution of each augmented code snippet results in the rendering of a new synthetic chart image. By repeating this iterative augmentation, we expand the initial dataset substantially, ultimately generating a dataset comprising over 200\,K unique synthetic chart image-code pairs from the original 13\,K seed charts.

The augmentation process is illustrated in Figure~\ref{fig:augmentation process}, and the detailed prompts utilized as part of the ChartGen pipeline are provided in the Appendix.

\section{ChartGen-200K Dataset}

\begin{figure}[htbp]
  \centering
  \includegraphics[width=1.0\linewidth]{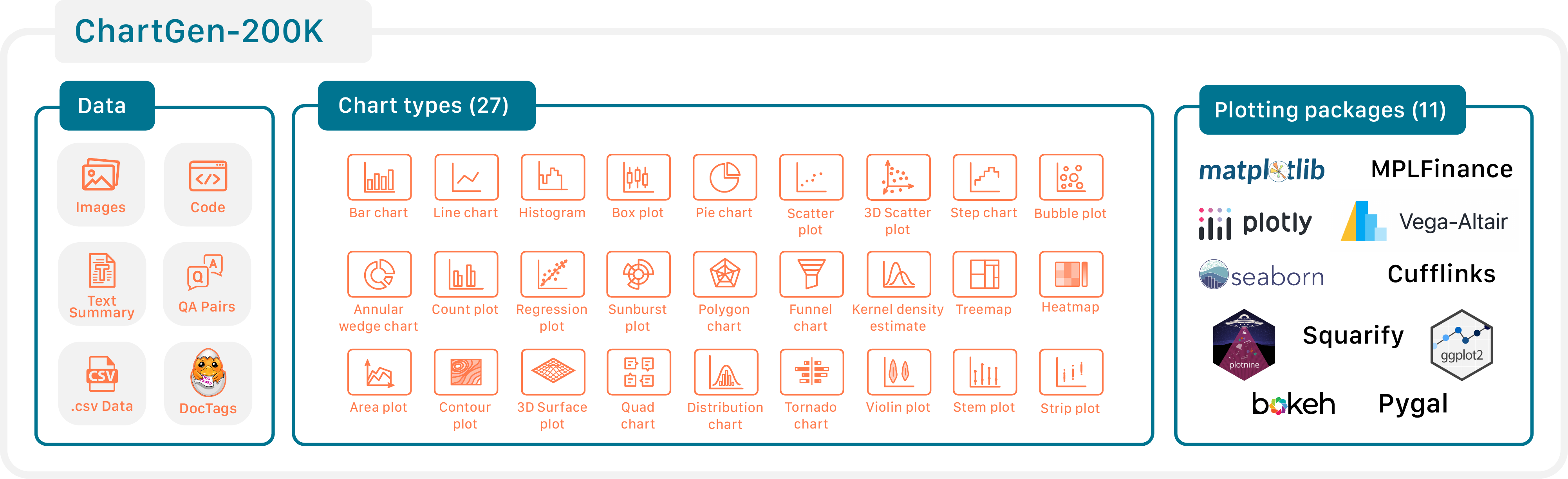} 
  \caption{An overview of the synthetic chart dataset obtained with ChartGen pipeline. The dataset includes 200K chart image-code-text-csv-doctags tuples, and contains 27 chart types, and the widest coverage of plotting packages to date}
  \label{fig:chartgen-200k}
\end{figure}

As outlined in Section \ref{subsec:data_gen}, the ChartGen pipeline results in a synthetic dataset composed of 222.5K unique chart image-code pairs. Each pair encompasses: (i) standard definition resolution JPG or PNG image files, and (ii) corresponding executable Python plotting scripts. Figure \ref{fig:chartgen-200k} presents an overview of the dataset, illustrating the coverage of chart types, plotting libraries, and included data modalities.

\subsection{Overview}

The ChartGen-200K dataset spans a diverse spectrum of 27 distinct chart types. Beyond traditional chart forms such as bar charts, line charts, scatter plots, and pie charts, the dataset includes more specialized visualizations like 3D plots, violin plots, heatmaps, and sunburst diagrams. This diverse coverage is achieved through a systematic augmentation strategy, wherein targeted prompts facilitate iterative transformations between different chart types  -- for example, converting one chart type into another, or transitioning from a simple chart to its three-dimensional equivalent.

An important hallmark of the ChartGen-200K dataset is its broad coverage of plotting packages, incorporating 11 distinct Python visualization libraries. These include both the well-established packages like \texttt{matplotlib}, \texttt{seaborn}, and \texttt{plotly}, as well as the other less commonly represented ones. This breadth enhances the dataset diversity by incorporating a wide variety of visualization layouts and distinct stylistic elements.

In addition, we maintain data quality through a two-part filtering procedure. First, we retain only executable Python plotting scripts, guaranteeing that all included code snippets can successfully render their corresponding visualizations. 
Additionally, duplicate charts are systematically detected and excluded based on identical image embeddings. This automated filtering mechanism ensures high functional consistency and dataset reliability.

\subsection{Supporting Chart Data}

Leveraging Python code as a foundational structured chart representation, we produce additional multimodal data components to further enhance the dataset’s utility in facilitating robust chart understanding. Specifically, besides the chart image-code pairs, the ChartGen-200K dataset additionally includes extracted CSV tabular data and DocTags to support chart-to-structure methods, as well as natural language text summaries and question-answer (QA) pairs. We provide the prompts used to generate the supporting chart data from the underlying chart code in the Appendix.

\paragraph{Extracted CSV Data}

Extracted CSV data provides the tabular representation underlying each chart visualization. This structured data directly supports tasks such as automated chart data retrieval, tabular data interpretation, and multimodal training scenarios where accurate numeric grounding is critical. We extract CSV data in a fully automated manner, prompting the Codestral-22B large language model to identify and extract the data objects referenced in the generated plotting code, ensuring consistent formatting and structure across the dataset.

\paragraph{DocTags}

DocTags is a compact, token-efficient representation that captures both semantic and structural attributes of documents, including charts, in a format suitable for LLMs/VLMs. It encodes chart types, spatial layout, and structural information in a way that supports seamless conversion to widely used formats such as HTML, Markdown, and JSON. We use Docling \citep{docling} to convert the previously generated CSV files into the DocTags format, enabling a structured and model-friendly representation of tabular data that retains compatibility while remaining easily exportable.


\paragraph{Natural Language Summaries}

Natural language summaries serve as critical input for multimodal tasks involving chart-to-text generation, summarization, and explanatory analysis. For each chart, we provide a paragraph-length natural language summary focusing on four key aspects: chart title, x-axis label, y-axis label, and data representations, in addition to an overall description of the chart layout. The summaries are generated by leveraging the text capabilities of \texttt{Codestral-22B}, using the previously generated plotting code as input. By feeding the model the underlying chart code -- rich with axis labels, series names, and metadata -- rather than a rendered image, we let it exploit these explicit semantics to craft more faithful and better grounded summaries.

\paragraph{Question-Answer (QA) Pairs}

To facilitate multimodal reasoning and evaluation, the ChartGen-200K dataset includes automatically generated question-answer pairs derived using the \texttt{Mixtral-8x22B} as well as off-the-shelf tools \citep{wood2024dataprepkit, digit}. The generation of QA pairs integrates multiple data modalities, including both the plotting code as well as the extracted CSV data and downstream-generated text summaries, to ensure comprehensive contextual grounding. These QA pairs span various complexity levels, including direct retrieval (e.g., "What is shown on the X-axis?"), as well as comparative reasoning (e.g., “Is the value for this category greater than, less than, or equal to the value for the other category overall?”).


\subsection{Comparisons to Seed Dataset} 

Below, we compare the seed dataset we use as input to the output dataset obtained with the ChartGen pipeline.
As shown in Table \ref{tab:dataset_comparison}, our method scales the size of the chart dataset by an order of magnitude -- from the 13\,K unique chart images in ChartQA \citep{masry2022chartqa} to 222.5\,K unique charts in the output dataset. In addition, while ChartQA focuses on standard bar, line, and pie charts, our dataset introduces a drastically broader array of chart types, reflecting diverse plotting libraries and code augmentations invoked by the ChartGen pipeline. To assess how the augmentations diversify the input charts, we report two measures as proxies for dataset diversity on both the seed set and the random equivalently-sized sample of our dataset:

\paragraph{Color Distribution (Shannon) Entropy.} To capture the variety in colors and their intensities across chart images, we compute their Shannon entropy, where higher entropy intuitively indicates a broader color palette. For each chart image, we compute a 256-bin RGB histogram, and estimate its Shannon entropy, $H = - \sum_{i=1}^{N} p_i \log_2 p_i$, where $p_i$ is the proportion of pixels in the $i$-th bin.

\paragraph{Semantic Embedding Spread} Besides the richness in purely visual appearance, we also explore the structural/ semantic diversity of the datasets, based on the chart image embeddings. We use a pretrained CLIP model \citep{radford2021clip} to embed each chart into a 512-dimensional latent space and measure the average pairwise (cosine) distance among the resulting feature vectors. Intuitively, charts that appear more distinct -- whether by their style or textual/ numerical content  -- will occupy more distant regions in the embedding space.

Table \ref{tab:dataset_comparison} summarizes our findings. Beyond the substantial gains in scale and chart-type coverage, the ChartGen pipeline produces an output dataset that is even more diverse than the real-world seed corpus, despite being entirely synthetic.
The increase in color entropy directly reflects the effects of the LLM-based code augmentations involving varying color schemes and plotting packages. In addition, the code augmentations diversify not only unique visual aspects such as chart orientations and color schemes, but also the structural/ semantic properties of the charts themselves, as indicated by the increase in average pairwise (cosine) distances between the chart image embeddings.

\begin{table}[hbt!]
\centering
\resizebox{0.7\textwidth}{!}{
\begin{tabular}{l|c|c|cc}
\toprule
 & \textbf{\# of Unique} & \textbf{\# of Chart} &
 \multicolumn{2}{c}{\textbf{Diversity Scores}} \\
\cmidrule(lr){4-5}
 & \textbf{Chart Images} & \textbf{Types} &
 \textbf{Color Entropy} & \textbf{Semantic Spread} \\
\midrule
\textbf{ChartQA} \citep{wei2024mchartqa}
 & 13\,K
 & 3
 & 2.03
 & 0.24
 \\
\textbf{ChartGen-200K (ours)}
 & \textbf{200\,K}
 & \textbf{27}
 & \textbf{2.06}
 & \textbf{0.38}
 \\
\bottomrule
\end{tabular}
}
\caption{Comparison of dataset scale and diversity metrics between the seed ChartQA dataset and our synthetic dataset produced by the ChartGen pipeline.}
\label{tab:dataset_comparison}
\end{table}

\subsection{Comparison with Existing Chart-Code Benchmarks \& Datasets}

Existing benchmarks and datasets involving chart understanding predominantly concentrate on tasks such as chart question answering, textual chart description, and summarization \citep{kahou2018figureqa, masry2022chartqa, kantharaj2022opencqa}, \citep{kantharaj2022charttotext, liu2024mmc, xu2024chartbench, zhu2025multichartqa, roberts2024scifibench}.  
In contrast, tasks that involve chart-to-code reconstruction remain underexplored. A key limitation in this area is the lack of large-scale, open-source datasets comprising chart images paired with their corresponding plotting code. 

Among the work most closely related to ours, \textit{Plot2Code} \citep{wu2024plot2code} benchmark offers 132 curated \texttt{matplotlib} charts spanning 6 chart types.  Every image is released together with (i) the exactly matching Python script extracted from the Matplotlib gallery and (ii) a GPT-4–generated natural-language instruction describing the visualization goals. 

\textit{ChartMimic} \citep{yang2025chartmimic} expands the scale to 1\,K samples and broadens the taxonomy to 22 chart types, while still focusing exclusively on \texttt{matplotlib}.  Each instance includes the web-crawled source chart, an “input chart’’ that the model must mimic, the corresponding human‐written Python code, and an instruction prompt.

Further increasing dataset scale, \textit{ChartX} \citep{xia2025chartxchartvlm} leverages GPT-4o to generate synthetic plotting codes and chart images, resulting in 6\,K validation and test samples. Crucially, each sample is multi-modal: besides the PNG and the plotting script, the authors release the CSV data table, textual chart summarization, and QA pairs.
This dataset moves beyond single-library corpora, including Python plotting libraries like \texttt{matplotlib}, \texttt{seaborn}, \texttt{plotly}, and \texttt{mplifinance}.
\texttt{}

More recently, \textit{ChartCoder} \citep{zhao2025chartcoder} provides a significant step-up in offered dataset size with approximately 160\,K chart image-code samples across 27 chart categories. The synthetic generation utilizes GPT-4o and relies on two plotting packages (\texttt{matplotlib} and \texttt{seaborn}).

Constructed via the ChartGen pipeline, our corpus represents, to our knowledge, the most extensive resource currently available for chart-to-code research. It comprises 222.5\,K samples including a held-out test split of 4.3 K chart image-code pairs, encompassesing 27 chart categories and 11 distinct Python plotting libraries.
Each chart image-code pair is accompanied by a comprehensive set of aligned modalities, including the underlying data table (CSV), DocTags structural markup, a concise natural-language summary, and automatically generated question–answer pairs -- facilitating the development and comprehensive evaluation of models capable of robust chart parsing.








\section{Chart Redrawing Evaluation}
\label{sec:benchmark}

To assess the ability of vision-language models to perform chart redrawing in a robust manner, we curate a dedicated evaluation set consisting of 4.3\,K chart image-code pairs drawn from our 222.5\,K corpus. Below, we formalize the task definition and describe the evaluation methodology.

\subsection{Task Definition}

The \textbf{chart redrawing} task evaluates the ability of a model to accurately recover the underlying plotting scripts from visual representations of charts.
Formally, given:
\begin{itemize}
\item A single chart image $\mathcal{I}$, originally generated by a Python plotting script $\mathcal{C}$,
\end{itemize}
the model is tasked with producing:
\begin{itemize}
\item A Python plotting script $\mathcal{C’}$ that closely matches the original script $\mathcal{C}$, thereby faithfully reconstructing the visual content and style of the chart depicted in $\mathcal{I}$.
\end{itemize}

\subsection{Evaluation Metrics}

We employ a two-pronged evaluation strategy, including comparisons of both the predicted code (direct textual comparison) and the resulting rendered images (visual comparison).  
We use GPT-4o as an automated judge in both scenarios, leveraging its language and vision capabilities for comprehensive assessments.

\paragraph{Chart Code Comparison}  In the first phase, we compare the Python code generated by the evaluated VLMs with the reference code (corresponding to the input chart image). In particular, we evaluate the models' performance using complementary criteria: (1) \emph{data fidelity} (i.e., whether the same underlying data distribution and values are being plotted) and (2) \emph{semantic/style consistency} (i.e., whether chart types, orientations, color schemes, and textual annotations match). We query GPT-4o with a prompt that shows both code snippets side by side and requests a pair of scores (please see Appendix for more details):

\begin{enumerate} \item \textbf{Data-based similarity score.}
A scalar measure (on a 0-1 scale) indicating how well the two code snippets represent the same data values, axes, and units of measurement.
\item \textbf{Semantic/style consistency score.}  
An additive scalar measure (on a 0-10 scale) reflecting whether the chart type (bar, line, scatter, etc.), orientation, labeling, legends, and color palettes are equivalent across the two snippets.
\end{enumerate}

GPT-4o provides both (a) a numerical score for each criterion and (b) a brief explanation justifying that score. We intentionally separate data fidelity from semantics/ style to capture differences in failure modes related to resolution-based visual complexities and numeric/ geometric reasoning versus those related to overall chart parsing.

\paragraph{Chart Image Comparison} In addition to comparing the code, we render the code generated by the models to produce a predicted (derendered) chart image. We then present both the input chart image and the model-generated chart image to GPT-4o (with vision capabilities) for a second round of comparison. We instruct it to produce a single aggregated similarity score (on a 0-10 scale), which additively accounts for a variety of qualitative criteria: alignment of chart types (bar, line, scatter, etc.), orientations, titles and axis labels, and overall styles and color schemes. Additionally, as before, we prompt GPT-4o to explain its reasoning, noting which specific elements match or differ.

By relying on image-level comparison, we aim to capture any discrepancies introduced by the chart image rendering process that might not be obvious from code alone.

\section{Results \& Discussion}

We evaluate six open-weight VLMs with total parameter size from 3B to 25.5 B: \texttt{Granite Vision} \citep{granitevisionteam2025}, \texttt{Phi-3.5-vision-instruct} \citep{abdin2024phi3}, \texttt{LLaVA-Next-Mistral} \citep{liu2024improvedbaselinesvisualinstruction}, \texttt{Molmo} \citep{deitke2024molmopixmo}, \texttt{Qwen-VL-Chat} \citep{bai2023qwenvl}, and \texttt{InternVL-2.5} \citep{chen2025expandingperformanceboundariesopensource} on our chart derendering benchmark, as described in Section \ref{sec:benchmark}. In Table~\ref{tab:multimodal_benchmarks}, we report the performance of each model according to four main criteria:

\begin{itemize}
    \item \textbf{Exec. Rate:} The fraction of generated Python code snippets that successfully execute without syntax or runtime errors.
    \item \textbf{Code Comparison (Data):} A scalar (range $0\text{-}1$) indicating correctness of the plotted data (axes values, bar heights, line coordinates, etc.), as judged by GPT-4o comparing the model’s code to the reference code.
    \item \textbf{Code Comparison (Semantics):} A score (range $0\text{-}10$) indicating how closely the generated script preserves chart type, orientation, labels, colors, and other high-level semantic or stylistic features, as judged by GPT-4o. 
    \item \textbf{Image Comparison:} A single GPT-4o-based rating ($0\text{-}10$) obtained by rendering the model-generated chart and comparing it visually to the ground-truth chart image.
\end{itemize}

\vspace{-0.5em}
\begin{table}[ht]
    \centering
    \resizebox{0.95\textwidth}{!}{
    \begin{tabular}{l l | c c c c}
        \toprule
        \textbf{Model} & \textbf{Params} & \textbf{Exec. Rate} & \textbf{Code Compar. (Data)} & \textbf{Code Compar. (Semantics)} & \textbf{Image Compar.}\\
        \midrule

        Granite-Vision-3.1 & 3B & 72.5\% & \textbf{0.58} & \textbf{5.83} & \textbf{7.48} \\

        Phi-3.5-vision-instruct & 4.15B & 66.9\% & 0.50 & 5.74 & 6.71 \\

        LLaVA-Next-Mistral & 7B & 67.1\% & 0.30 & 4.76 & 5.42 \\

        Molmo & 7B & 20.6\% & 0.38 & 4.92 & 6.14 \\
        
        Qwen-VL-Chat & 9.6B & 74.5\% & 0.50 & 5.48 & 6.67 \\
                
        InternVL-2.5 & 26B  & \textbf{86.0}\% & 0.45 & 5.54 & 7.07 \\

        \bottomrule
    \end{tabular}
    }
    \caption{Overall performance comparison of six vision-language models on our chart derendering benchmark. \emph{Exec. Rate} is the fraction of generated Python code snippets that run without errors. \emph{Code Compar. (Data)} ranges in $[0,1]$, while \emph{Code Compar. (Semantics)} and \emph{Image Compar.} range in $[0,10]$.}
    \label{tab:multimodal_benchmarks}
\end{table}

\paragraph{Chart derendering remains a challenge for open-weight vision-language models.} The quantitative evaluation, as outlined in Table \ref{tab:multimodal_benchmarks}, reveals that accurately interpreting and reproducing charts into executable and semantically correct Python code remains a challenging task. Although execution rates are moderately high for the most part, indicating models’ ability to produce syntactically valid code, the correctness of the plotted data and chart semantics largely lags behind. This discrepancy suggests that reliably capturing numerical values and relationships (e.g., axes scaling, bar heights, or line coordinates) and preserving stylistic elements (such as chart orientation, color schemes, and labeling conventions) are indeed nontrivial for current state-of-the-art open-weight models.  Moreover, the gap between syntactically valid code and faithful reproduction of the intended visual representation, as indicated by the image comparison task, underscores the multifaceted nature of chart redrawing, which requires precise data handling, stylistic consistency, and visual alignment. These results highlight that extracting structured, semantically accurate, and visually faithful chart representations from images continues to pose substantial challenges for current open-weight VLMs.

\section{Related Works}

\paragraph{Chart Reasoning} 
Early work on chart understanding largely focused on \emph{question-answering} tasks, where a model is given a chart image and asked to answer queries of varying complexity about its content (e.g., \citep{kahou2018figureqa, masry2022chartqa, kantharaj2022opencqa, zhu2025multichartqa}). Benchmark datasets such as \emph{FigureQA} \citep{kahou2018figureqa} and \emph{ChartQA} \citep{masry2022chartqa} spurred interest in models capable of parsing visual elements (e.g., axes labels, legend entries) and inferring underlying data relationships. Subsequent efforts extended this to open-ended summaries and text generation from charts \citep{kantharaj2022charttotext, roberts2024scifibench}, as well as multi-modal tasks involving both textual and tabular information \citep{xu2024chartbench, liu2024mmc}. In contrast, our approach shifts attention from answering factoid or summary-style questions to \emph{chart reconstruction} -- generating a complete chart representation in the form of plotting code that captures both data details and stylistic properties.  Importantly, code serves as a comprehensive and structured representation of charts, enabling downstream chart data generation, such as creating textual summaries or generating question-answer pairs directly from the code. In addition, this “chart-to-code” focus provides a more granular assessment of model grounding and reasoning abilities, facilitating more comprehensive evaluations than question answering alone.

\paragraph{Vision-Language Models} Recent advancements in vision-language models (VLMs) \citep{bai2023qwenvl, li2024llavaonevisioneasyvisualtask, agrawal2024pixtral12b, granitevisionteam2025, meta2025llama4} have significantly extended the multimodal capabilities of large language models (LLMs) \citep{brown2020languagemodelsfewshotlearners}. However, existing evaluation benchmarks primarily focus on tasks like captioning or visual question answering (VQA), which may not fully capture the complexities of chart interpretation and data visualizations more broadly, or real-world applicability \citep{li2024surveybenchmarksmultimodallarge}. To better assess multimodal reasoning and grounding, we propose evaluating VLMs through a \emph{chart redrawing} task -- an approach that requires detailed visual parsing and structured representation of chart data and semantics, providing a more rigorous test of their capabilities in interpreting complex visual information.

\paragraph{Multimodal Code Generation}
Early code-generation benchmarks primarily concentrate on text-based or single-modal prompts \citep{chen2021evaluating, austin2021programsynthesis, lai2022ds1000}. 
With the increasing sophistication of VLMs, there has been a growing interest in \textit{image-driven} coding tasks. Among the image-to-code tasks, \citep{si2025design2code, soselia2023learningui} evaluate models on their ability to reverse-engineer HTML code directly from UI screenshots. On the other hand, \citet{yang2025scalingimageunderstanding} leverages the coding abilities of LLMs to generating multimodal data. Focusing explicitly on chart generation, recent works such as \emph{Plot2Code} \citep{wu2024plot2code}, \emph{ChartMimic} \citep{yang2025chartmimic}, \emph{ChartX} \citep{xia2025chartxchartvlm}, and \emph{ChartCoder} \citep{zhao2025chartcoder} have explored small- to mid-scale datasets dedicated to chart-to-code tasks. However, these resources typically span a few hundred to a few thousand chart image-code samples (with the exception of ChartCoder encompassing a 160\,K samples), often restricted to limited plotting libraries. Addressing this gap, our work introduces 222.5\,K-sample multimodal synthetic chart dataset designed to facilitate robust chart reasoning and systematic evaluation of chart code generation. By emphasizing diverse chart styles and various plotting packages, ChartGen-200K paves the way for more comprehensive and generalizable chart reasoning research.

\section{Conclusion}

We introduce ChartGen, a two-stage, fully–automated pipeline that first reconstructs real-world charts into code with a vision–language model and then diversifies the generated code with a code-centric LLM.  Applied to 13 K seed charts, ChartGen synthesises 222.5\,K image-code pairs that span 27 chart types and 11 plotting libraries, each accompanied by tabular CSV data, DocTags, natural-language summaries, and QA pairs -- an order-of-magnitude increase over prior resources.  From this corpus we distill a 4.3\,K held-out chart redrawing evaluation set and propose a GPT-4o–based protocol that jointly scores the generated code fidelity and rendered-image similarity, revealing a wide performance gap across six open-weight VLMs (best data-fidelity = 0.58/1; best image-similarity = 7.48/10).  These results highlight that chart-to-code reconstruction remains an unsolved problem and underscore the need for richer multimodal chart datasets.  While ChartGen substantially expands the multimodal chart landscape, it unavoidably inherits representational biases from its underlying vision-language and code models. Future work will tackle these biases in tandem with efforts toward even larger-scale generation and richer reasoning over chart semantics. Beyond scale, ChartGen’s use of code as a canonical representation enables efficient generation of supporting chart data such as summaries or question-answer pairs, making it a useful resource for future research in training and evaluating chart understanding and vision-conditioned code generation more broadly. 


\section{Broader Impacts}

This line of work has significant positive implications, particularly related to facilitating robust automated chart understanding and supplying the large-scale supervision needed to train next-generation multimodal AI. By translating complex figures into editable code and accessible descriptions, ChartGen can potentially benefit applications in accessibility, education, and data-driven decision-making. However, reliance on synthetic data carries risks of propagating inaccuracies if not adequately validated, highlighting the importance of critical use and continuous improvements in automated verification as well as human oversight.

\section*{Acknowledgments}
This work was partially funded by the MIT–IBM Watson AI Lab.  
We also thank Bernhard Paus Graesdal for his invaluable contributions to the figure design.

\newpage

\bibliographystyle{plainnat}
\bibliography{references}

\newpage

\appendix

\section{Generation Configurations and Prompt Templates}

In this section, we specify the model and compute configurations, and the exact prompts used in Section 2, Section 3.2, and Section 4.

\subsection{The ChartGen Pipeline}

For the first stage of the ChartGen pipeline, we employ the vision–language model \texttt{phi-3.5-vision-instruct}, with maximum generation length set to $2000$ new tokens, and temperature set to $\tau =0.0$ to obtain deterministic outputs and increase the likelihood of faithful chart reconstruction.
For chart code augmentation we rely on \texttt{Codestral-22B-v0.1}, configured to output at most $1200$ tokens with sampling disabled.
Creating the synthetic dataset comprised of 222.5\,K chart image-code pairs consumed approximately $1000$ GPU hours on NVIDIA A100 (80 GB) hardware.

\textbf{Stage 1: VLM-based Chart-to-Code Reconstruction}

\begin{figure}[ht]
  \centering
  \includegraphics[width=0.9\textwidth]{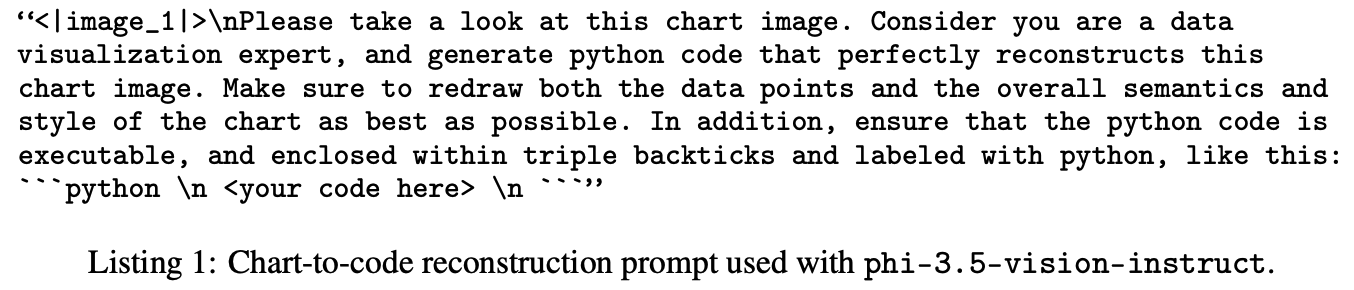}
\end{figure}

\textbf{Stage 2: LLM-based Code Augmentation}

\begin{figure}[ht]
  \centering
  \includegraphics[width=0.9\textwidth]{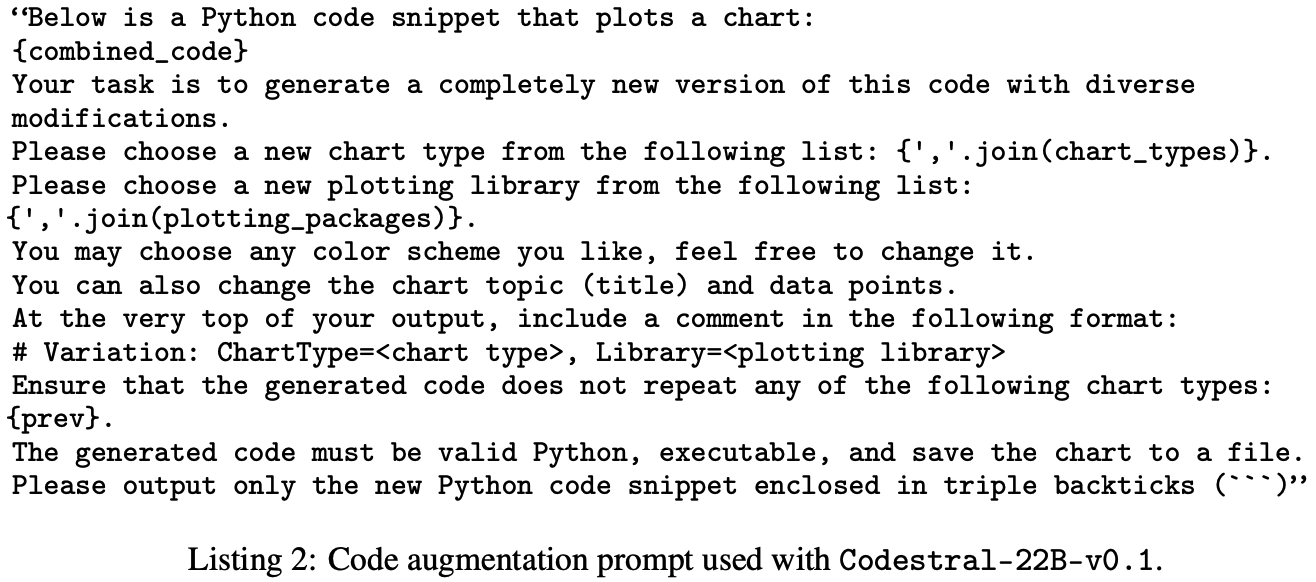}
\end{figure}

\subsection{Supporting Chart Data}

In addition to chart image-code pairs generated using the ChartGen pipeline, we also create supporting chart data, including CSV data and text summaries generated using \texttt{Codestral-22B-v0.1}, and question-answer pairs generated using \texttt{Mixtral-8x22B}. Below, we specify the prompts and configurations used. 

\subsubsection*{Extracted CSV Data}

For every chart code generated by the ChartGen pipeline, we extract a clean tabular representation of the plotted data.  
We employ \texttt{Codestral-22B-v0.1} with deterministic decoding and maximum generation length of
$512$ tokens.  
The extraction pass over the full corpus consumed roughly $250$ GPU-hours on NVIDIA A100 (80 GB) hardware.

\begin{figure}[ht]
  \centering
  \includegraphics[width=0.9\textwidth]{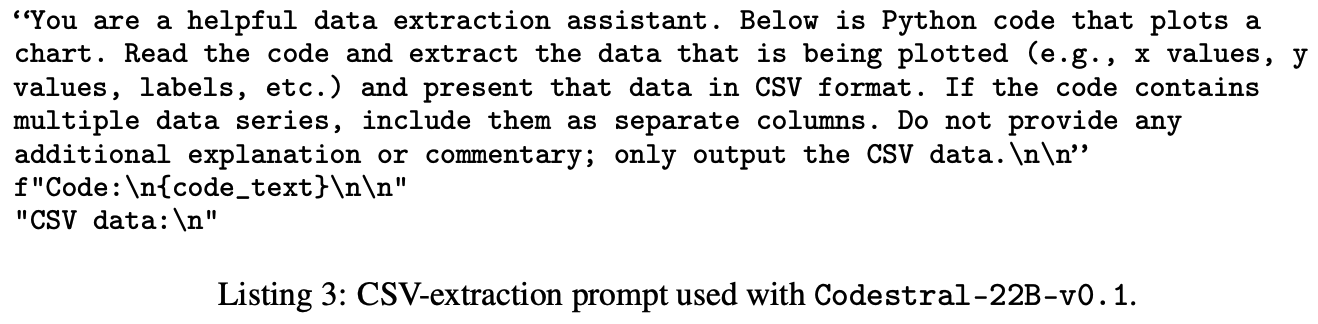}
\end{figure}

\subsubsection*{Natural Language Summaries}

Using the chart code generated with the ChartGen pipeline, we automatically produce text summaries of what a viewer would see. We use \texttt{Codestral-22B-v0.1} with a mildly stochastic decoding setup -- temperature $\tau = 0.2$ and maximum generation length of $1000$ tokens. Generating summaries for the full corpus consumed roughly 500 GPU-hours on NVIDIA A100 (80 GB) hardware.

\begin{figure}[ht!]
  \centering
  \includegraphics[width=1.0\textwidth]{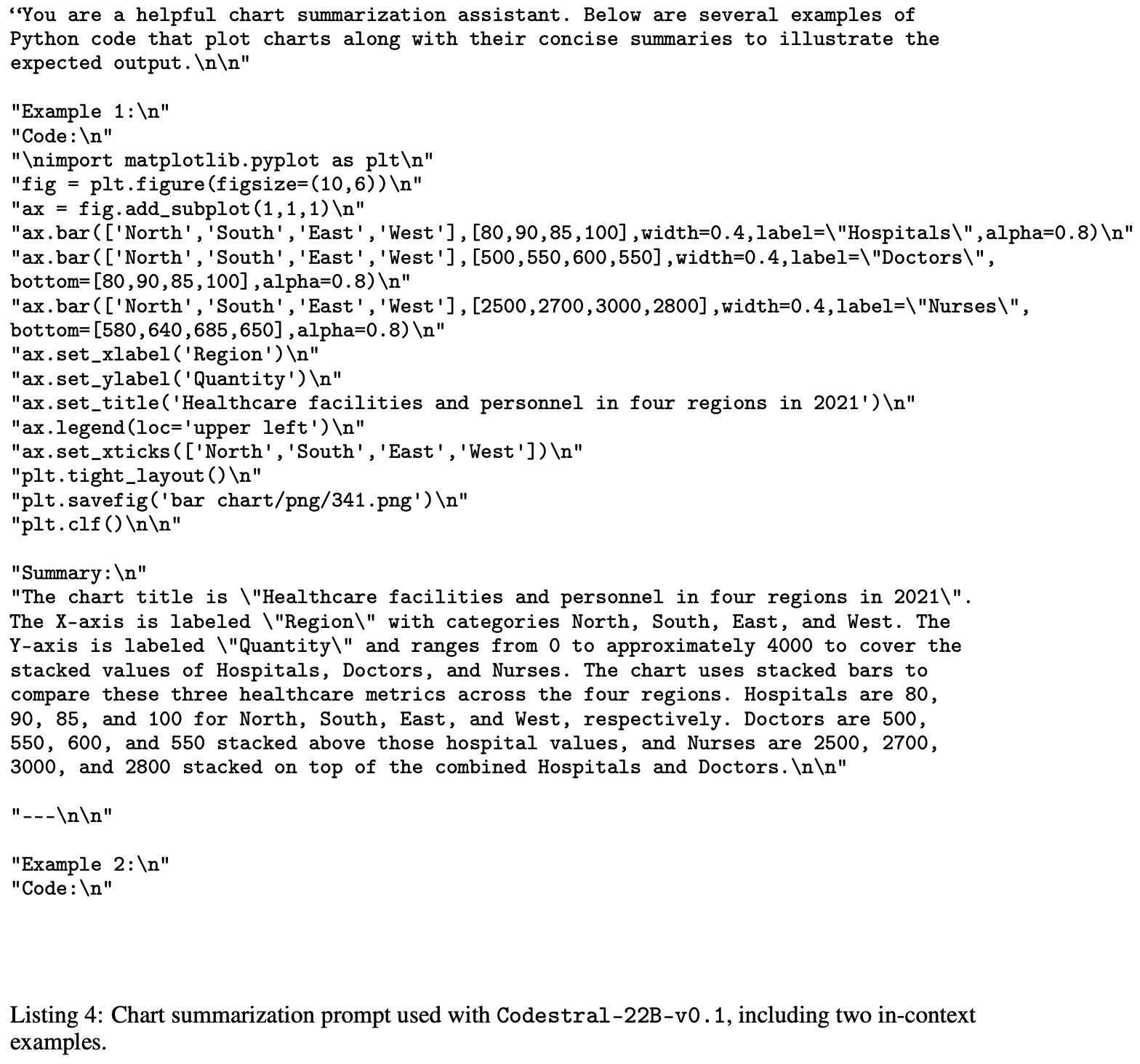}
\end{figure}

\begin{figure}[ht!]
\hspace{-1.5cm}
  \centering
  \includegraphics[width=0.9\textwidth]{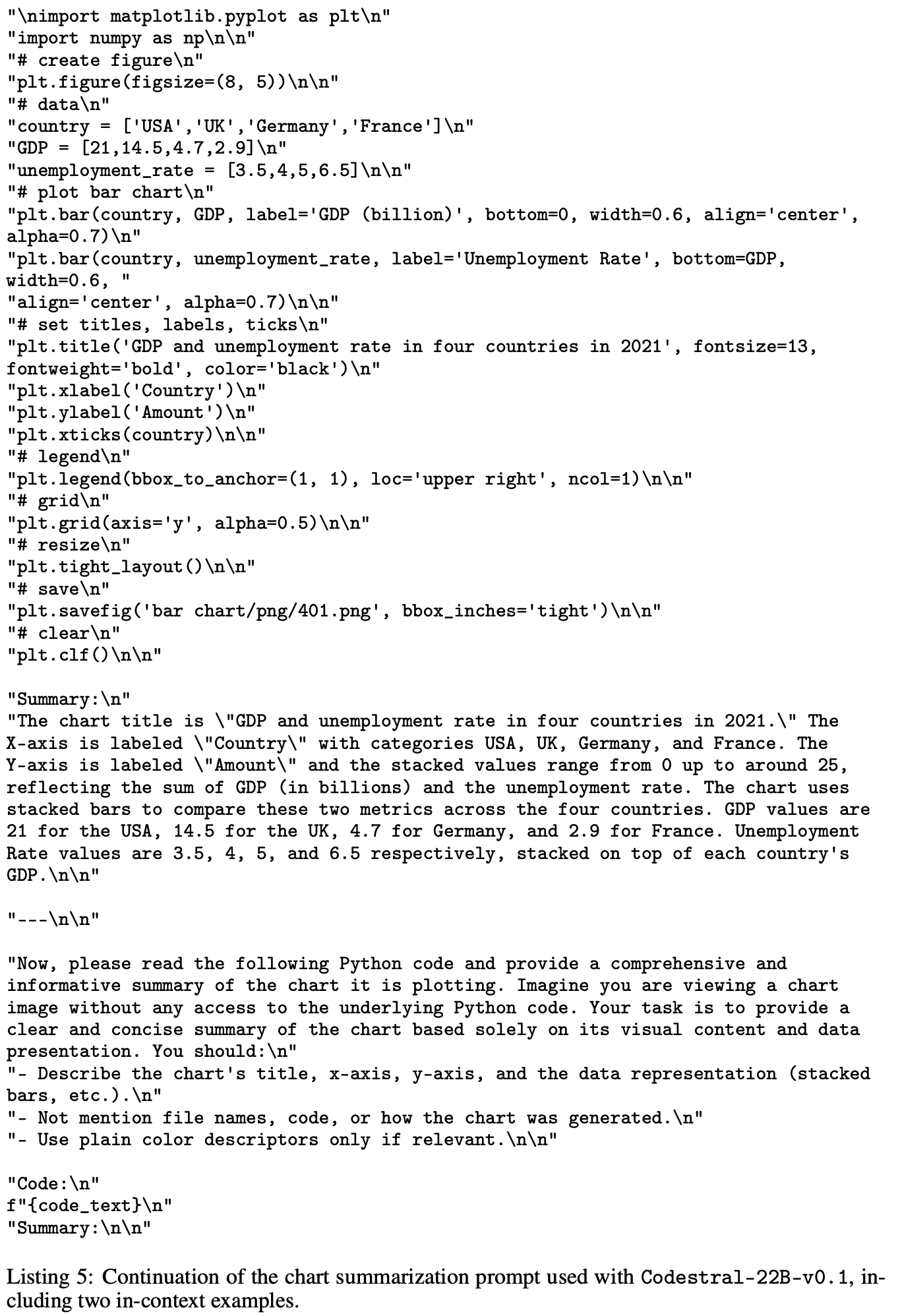}
\end{figure} 

\subsubsection*{Question-Answer Pairs}

Using the generated chart code, CSV data, and text summaries as context, we also generate  QA pairs using \texttt{Mixtral-8x22B}, consuming roughly 600 GPU-hours on NVIDIA A100 (80 GB) hardware.

\begin{figure}[ht]
  \centering
  \includegraphics[width=0.87\textwidth]{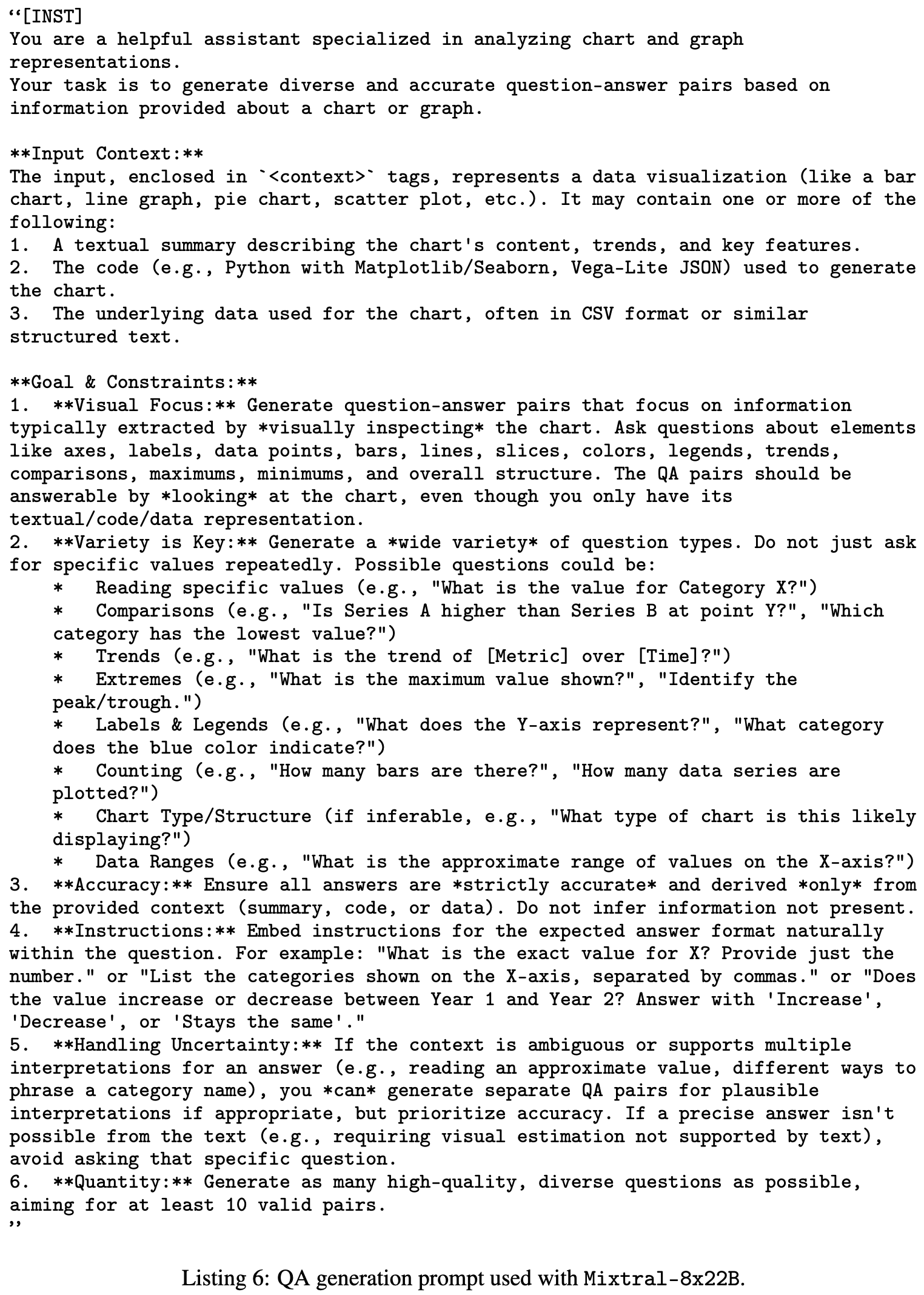}
\end{figure}

\begin{figure}[ht]
  \centering
  \includegraphics[width=0.89\textwidth]{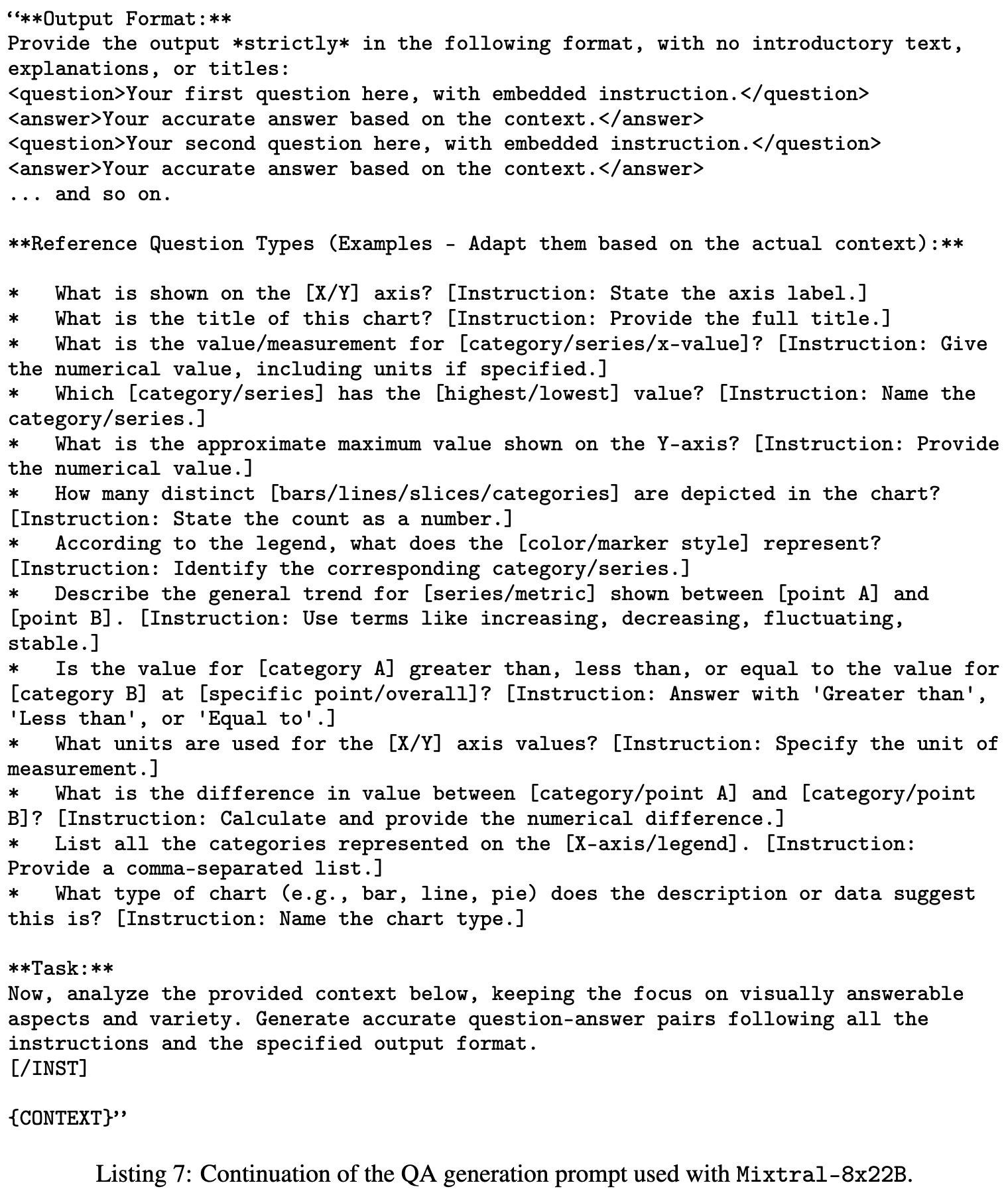}
\end{figure}

\subsection{Chart Redrawing Evaluation}

For all of the six vision-language models that were evaluated on our chart redrawing evaluation set, we configure the maximum generation length to 1200 tokens and the temperature $\tau=0$ for greedy decoding. All inference was performed on a single A100 80GB GPU. We report the prompt used with each model below.

\begin{figure}[ht]
  \centering
  \includegraphics[width=0.9\textwidth]{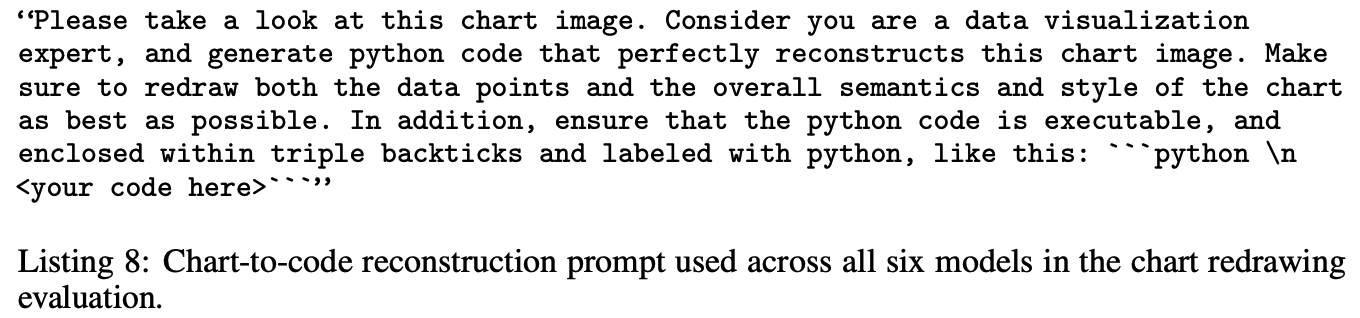}
\end{figure}

\newpage
\subsubsection*{GPT-4o-based Evaluation}

To evaluate the performance of the six models on the chart redrawing task, we use GPT-4o as an automated judge. All calls are made through the OpenAI API with deterministic decoding. Below we outline the prompts used for our two evaluation metrics -- chart code comparison and chart image comparison.

\begin{figure}[ht]
\hspace{-1cm}
  \centering
  \includegraphics[width=0.815\textwidth]{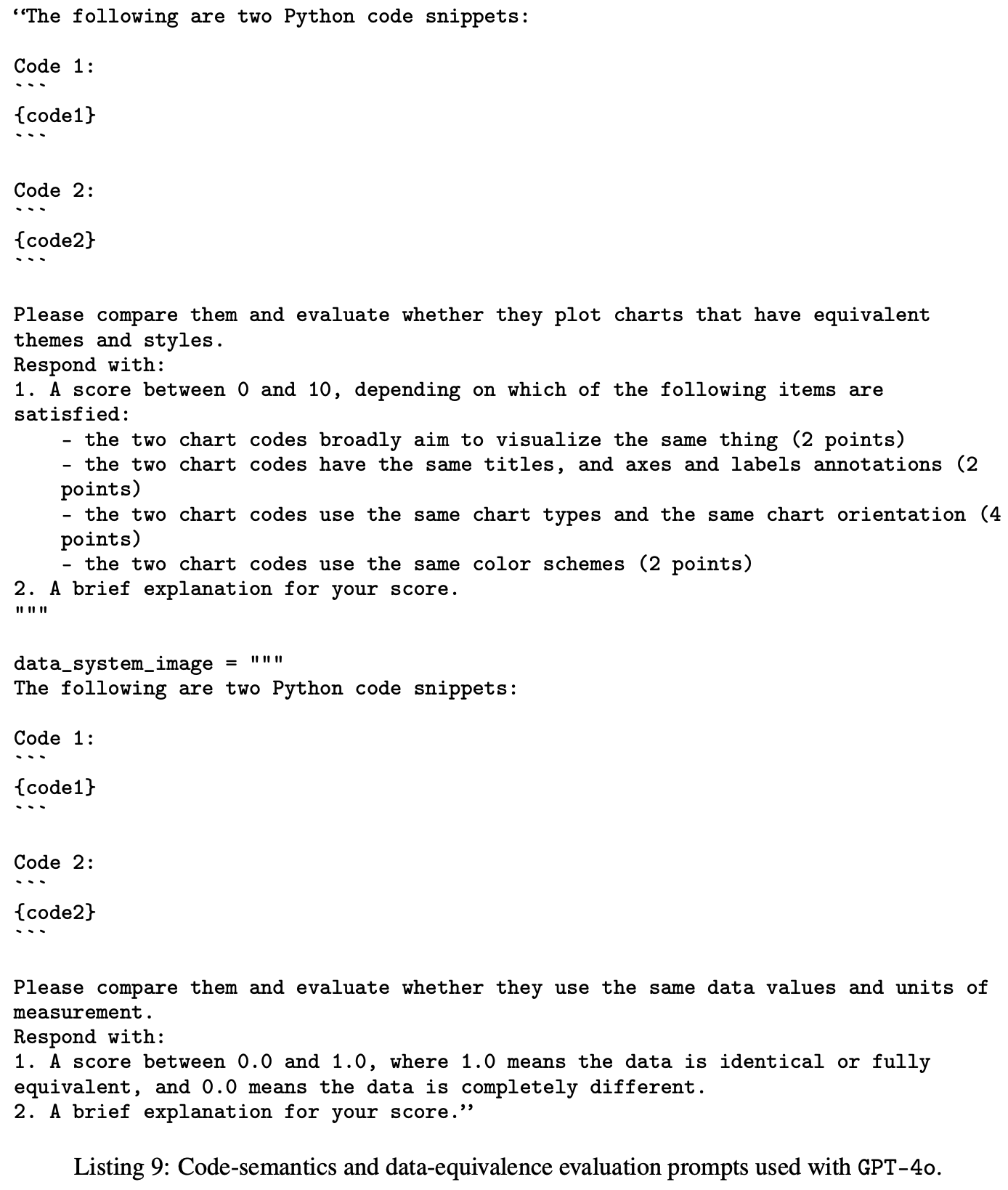}
\end{figure}

-
\begin{figure}[ht]
\hspace{-2mm}
  \centering
  \includegraphics[width=0.85\textwidth]{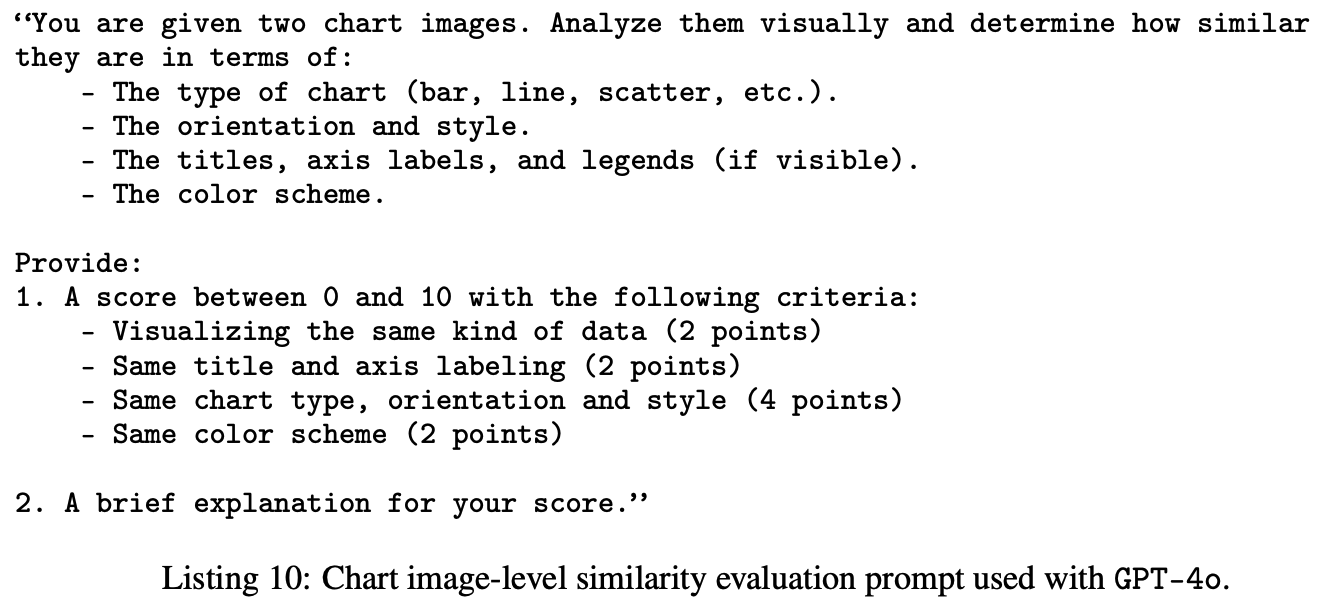}
\end{figure}

\end{document}